\documentclass[a4paper,11pt]{article}
\usepackage{jinstpub} % for details on the use of the package, please see the JINST-author-manual
\usepackage{lineno}
\usepackage[export]{adjustbox}

\title{\boldmath {The potential applications of muography to revealing sea shipwrecks}}

\author{Anzori~Sh.~Georgadze}
\affiliation{Kyiv Institute for Nuclear Research, Prospekt Nauky 47, 03680, Kyiv, Ukraine}

\emailAdd{a.sh.georgadze@gmail.com}
\abstract{
Muon imaging, a non-invasive technique that utilizes naturally occurring cosmic muons, has emerged as a promising tool for exploring underwater objects, including shipwrecks. This study investigates the potential of muon radiography to examine the contents of wrecked ships in the Baltic Sea and other marine environments. These wrecks often pose significant environmental risks due to hazardous contents such as explosives and crude oil, making their detection and monitoring critical for environmental and safety considerations. Accurate modeling and imaging of such wrecks are therefore essential for assessing potential dangers and mitigating environmental impacts.

To model the underwater muon flux in this study, an approach similar to that used in underground muon experiments was adopted. Cosmic muons were propagated through seawater using GEANT4, modeling their energy loss as they traverse water. The muons' kinematics and energy depositions were recorded during this process. The resulting distributions were tabulated and used as input for the primary particle generation module. This method enabled the generation of a realistic underwater muon flux at the desired depth above the shipwreck, allowing us to simulate various shipwreck filling scenarios.

Assuming an exposure time of one week for a wreck located at 50 meters depth, our simulations demonstrate that muon imaging can sufficiently resolve density contrasts to distinguish between water, oil, and high-density materials. These results demonstrate the feasibility of muon radiography as a practical tool for underwater hazard assessment and shipwreck investigation.
}


\arxivnumber{2509.00164} % Only if you have one

\begin{document}
\maketitle
\flushbottom

%\textcolor{red}{
\section{Introduction}
The deep sea holds countless sunken vessels, some of which may contain hazardous materials such as oil or explosives, making direct investigation risky for both safety and the preservation of the wreck.
A promising, non-invasive solution comes from particle physics: muon imaging. Also known as muon tomography or radiography, this technique exploits naturally occurring cosmic-ray muons, highly penetrating particles that can reveal the internal structure of large, dense objects. Initially developed for applications such as scanning pyramids and monitoring nuclear reactors, muon imaging has since expanded to geophysics, archaeology, civil engineering and border security~\cite{tanaka2007high, lesparre2012density, morishima2017discovery, das2022muography, saracino2017imaging, nishiyama2017first, tripathy2021numerical, thompson2020muon, saracino2019applications, mao2023muon, georgadze2025muon}. Muon imaging has also been explored for underwater applications, with pilot studies demonstrating its considerable potential. Researchers at the University of Tokyo have successfully employed muon radiography to image underwater volcanic structures, showcasing its feasibility in marine environments. This technique involves deploying muon detectors beneath the water to visualize subsurface geological features. A notable example is the Tokyo Bay Seafloor Hyper-Kilometric Submarine Deep Detector (TS-HKMSDD), which is the world’s first under-seafloor muon detector array. Installed beneath the Tokyo Bay seafloor, this system collects muon data to monitor tidal variations and other subsurface phenomena\cite{MAGMA2023}. Additionally, the University of Tokyo's International Muography Research Organization has developed the wireless muometric navigation system (MuWNS), which leverages cosmic-ray muons to enable navigation in underground environments where GPS signals are unavailable\cite{tanaka2023first}.  
These advancements highlight the promising potential of muon radiography as a non-invasive technique for underwater exploration and monitoring. 
In this context, its application to submerged environments offers exciting possibilities. By utilizing muon imaging to examine shipwrecks, researchers can analyze internal structures and contents without the need for excavation. This approach not only aids in the preservation of cultural heritage but also opens new avenues for uncovering historical and scientific insights beneath the waves.

The Baltic Sea has served as a strategic maritime corridor for centuries, linking Northern and Eastern Europe through trade, exploration, and conflict. This long history has left behind a dense concentration of shipwrecks, with estimates ranging from 10,000 to over 100,000 vessels lost beneath its waters \cite{Hac2024}. These wrecks span diverse historical periods, from the \textit{Middle Ages} and \textit{Hanseatic League} to both \textit{World Wars}, reflecting the region's pivotal role in European naval and commercial history \cite{westerdahl1992}.

Underwater shipwrecks pose unique challenges for radiographic detection, particularly when hazardous materials are involved. Crude oil and diesel fuel, stored in tanks, show moderate muon attenuation with relatively little scattering, making them relatively straightforward to detect. Explosives cause stronger attenuation combined with moderate scattering, complicating imaging. 
Steel-cased bombs, by contrast, create significant scattering and attenuation due to their high density. Voids or collapsed compartments appear as bright, low-density anomalies. Well-known examples of wrecks with hazardous cargo include the ammunition ship \textit{S.S. Richard Montgomery}, the crude oil tanker \textit{Montebello}, and the \textit{Lusitania}, long rumored to have carried munitions. In the Baltic region, many wrecks are suspected to contain either crude oil or chemical munitions, but their internal condition remains uncertain.

Crude oil leakage from shipwrecks poses severe environmental and economic risks. Released crude oil contaminates seawater, impairs reproduction and respiration in marine organisms, and damages sensitive ecosystems such as reefs, seagrass beds, and wetlands. These effects cascade into fisheries and tourism, creating long-term economic losses. Cleanup operations require specialized equipment, trained personnel, and prolonged efforts, often costing millions to billions of dollars. Preventing leaks and implementing effective monitoring and response strategies are therefore essential for safeguarding marine ecosystems and local economies.

\textbf{Traditional methods for shipwreck exploration}.

Shipwreck exploration traditionally relies on a combination of archival research, diver surveys, controlled underwater excavations, geophysical surveys, robotic platforms, and non-invasive analyses. Historical documents and naval records guide the initial identification of candidate sites, while divers, when conditions and depth allow, provide direct inspection and sampling. 

Geophysical methods such as side-scan sonar, multibeam echo sounding, magnetometry, and sub-bottom profiling are widely used to map wreck structures and surrounding sediments. Remotely Operated Vehicles (ROVs) and Autonomous Underwater Vehicles (AUVs), equipped with cameras and sonar, extend surveys to greater depths or hazardous environments. Photogrammetry and 3D modeling techniques enable detailed reconstructions of wreck morphology, while non-invasive laboratory analyses, including wood identification, dendrochronology, isotopic analysis, and corrosion studies, provide valuable insights into age, origin, and preservation state.

Despite their successes, these techniques have several shortcomings. Optical and acoustic methods primarily reveal external structures and are limited in their ability to probe internal compartments or distinguish materials of different densities. Sub-bottom profilers can penetrate sediments but only resolve large-scale features, while magnetometry detects metallic objects but cannot unambiguously characterize their composition or content. Diver surveys are restricted to relatively shallow depths, and underwater excavations are invasive, expensive, and risk damaging fragile cultural heritage. Furthermore, turbidity, biofouling, sediment cover, and corrosion often obscure or distort measurements, reducing accuracy. 

As a result, traditional approaches can document external morphology and site context, but they remain limited when it comes to non-invasively mapping the internal structure, cargo, or hazardous materials of wrecks. This motivates the exploration of complementary approaches such as muon radiography, which can image density distributions through thick sediment and metallic hulls without physical disturbance.

\textbf{Underwater muon imaging for shipwreck exploration}.

The method is particularly suited to scanning large dense objects. Muon attenuation signatures when passing through this objects can reveal crude oil and fuel tanks, explosives, or steel-cased munitions~\cite{helcom2025, szpiech2024baltic,  cimmino2021principles}. Feasibility has already been demonstrated in Tokyo Bay, where detectors imaged through tens of meters of water and sediment \cite{tanaka2021tokyobay}. Similar deployments could be adapted for Baltic wrecks, generating either 2D transmission maps or full 3D tomographic reconstructions. Advances in compact scintillator arrays, readout electronics, and pressure housings make such systems increasingly feasible \cite{pezzotti2025new}, while simulations with \texttt{GEANT4} provide guidance for geometry optimization and exposure time.  

\section{Methods and tools}

\subsection{Monte Carlo simulations}

\begin{figure}[t]
\centering
\includegraphics[width=0.54\textwidth]{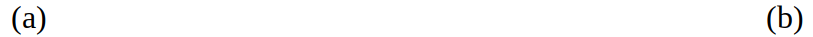}
\includegraphics[width=0.49\textwidth]{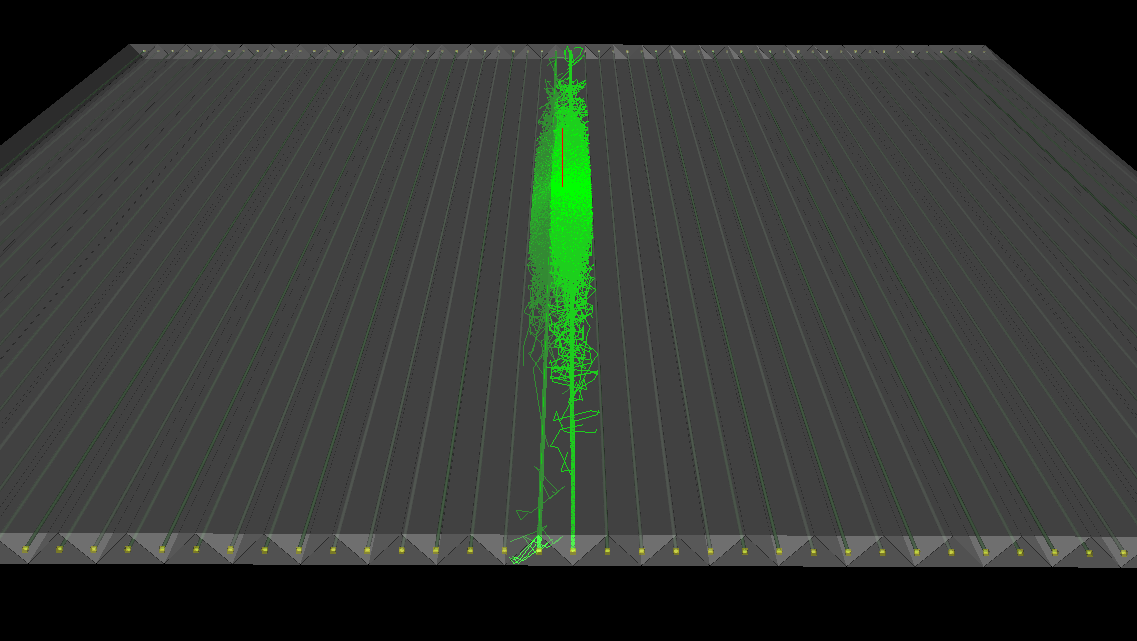}
\includegraphics[width=0.49\textwidth]{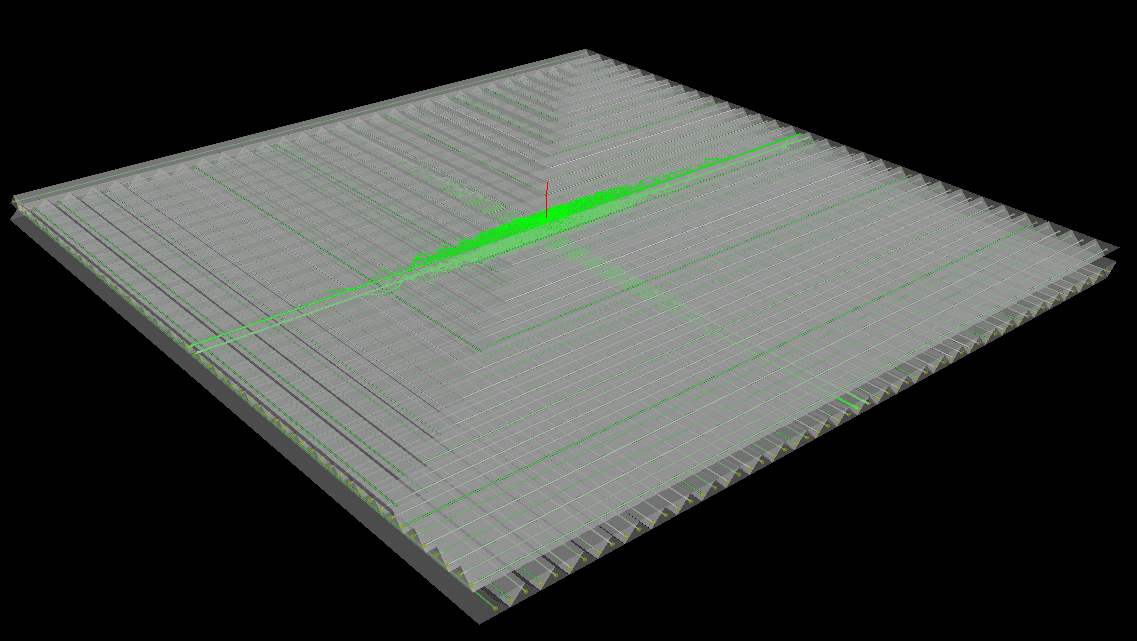}
\caption{(a) GEANT4 simulation of optical photon propagation in a single layer of stacked triangular bars is shown. (b) Two orthogonally oriented layers form a $1 \times 1~\mathrm{m}^2$ superlayer used for determining the $x$–$y$ coordinates of the muon hit position.}
\label{fig:f1op}
\end{figure}

In GEANT4~\cite{Agostinelli2003GEANT4}, the detector is modeled as a set of three planes, each measuring \(1 \times 1\, \mathrm{m}^2\), with a typical vertical spacing of 0.5 m. This configuration functions as a muon telescope for tracking muons and reconstructing internal density via muon radiography.
Each detector plane consists of two adjacent modules, totaling 64 plastic scintillator bars with triangular cross-sections and central holes for wavelength-shifting (WLS) fibers. The triangular geometry enables compact, crack-free arrays, while combining signals from adjacent bars improves spatial resolution through weighted position averaging.
WLS fibers collect scintillation light and transmit it to silicon photomultipliers (SiPMs). Each plane provides one spatial coordinate (\textit{x} or \textit{y}) of the muon hit. Two planes with orthogonal bars form an \textit{x}-\textit{y} station, and the complete muon telescope consists of three such stations (Figure~\ref{fig:f1op}).
Muon trajectories are reconstructed by fitting straight lines through the recorded (\textit{x},\textit{y},\textit{z}) hit positions across the three stations. The three-layer redundancy enhances track accuracy, reduces accidental triggers from SiPM noise and natural radioactivity.

In marine environments, detectors are positioned near wrecks to detect muons passing through structures, enabling noninvasive 3D imaging of internal components. To enhance detection efficiency, the detector is inclined at a 10-degree angle relative to the wreck.

\subsubsection{Modeling optical photons transport}
We have developed a customizable simulation code for modeling optical photon transport in organic scintillators based on GEANT4. The code allows for the implementation of various geometries, materials, and reflective surfaces.

The scintillator bar is 1000 mm long, have a triangular cross-section (as shown in Figure~\ref{fig:f1reco}(a)) with a base width of \(33~\mathrm{mm}\), a height of \(17~\mathrm{mm}\), and an extrusion hole of \(2~\mathrm{mm}\) in diameter to accommodate the wavelength-shifting (WLS) fiber.
The EJ-200 plastic scintillator~\cite{eljen} was selected as the base material. A scintillation yield of \(10{,}000~\text{photons/MeV}\) was assumed to benchmark the optical model. 

Extruded plastic scintillator bars are co-extruded with a white outer coating composed of 15\% of titanium dioxide (TiO$_2$) mixed with polystyrene. This coating functions as a diffusive internal reflector for scintillation light. By reflecting and directing scintillation photons toward the embedded wavelength-shifting fibers within the scintillator core, the TiO$_2$ coating enhances light collection efficiency. The co-extrusion process for applying this reflective layer is more cost-effective than alternative methods, such as wrapping the scintillator with reflective materials, and it provides a more uniform, air-free coating. The light attenuation length of scintillator bar was \textit{L}= 30 cm~\cite{pla2005extruded}.

The scintillator strips are read out by 1.5 mm diameter Kuraray multiclad WLS fibers, 11-Y (S35 type)~\cite{kuraray}. The SiPMs detect scintillation light from both ends of the WLS fibers. To reduce light losses, an optical coupler was applied in a 2.0 mm diameter hole, which formed the gap between the fiber and the scintillator bars.
The scintillator bars have an index of refraction \( n_{\text{sc}} = 1.59 \), while the WLS fibers have indices of refraction \( n_{\text{WLS}} = 1.59 \), 1.49, and 1.42 for the fiber core, inner cladding, and outer cladding, respectively. The silicone optical grease used as the coupler in the gap had a refractive index of \( n_{\text{gap}} = 1.50 \), which optimizes light transmission and reflection properties.
Hamamatsu S13360-2050VE devices~\cite{hamamatsus13360} with a 2 × 2 mm$^2$ photosensitive area is used to detect optical photons. The SiPMs are modeled as a Si material. 
To improve light detection, SiPMs are coupled with WLS fiber with an Silicone Optical Grease BC-631~\cite{luxium} with a refractive index of 1.465. 
The light detection efficiency is determined by weighting the SiPM PDE with the Kuraray WLS fiber emission spectra, which gives a value of $\approx$40\%.

In this work, the \texttt{glisur} model is employed the, which accounts for surface roughness (polishing degree) through a roughness parameter in the range 0.5–0.9.
The surfaces are modeled as  \texttt{dielectric{\_}dielectric}, co-extruded coating was treated as \texttt{dielectric{\_}metal} surface, where photons are reflected with a probability corresponding to the set reflectivity, non-reflected photons are absorbed.

\subsubsection{Position resolution estimation} 

\begin{figure}[b]
\centering
\includegraphics[width=0.54\textwidth]{figures/ab.png}
\includegraphics[width=0.46\textwidth]{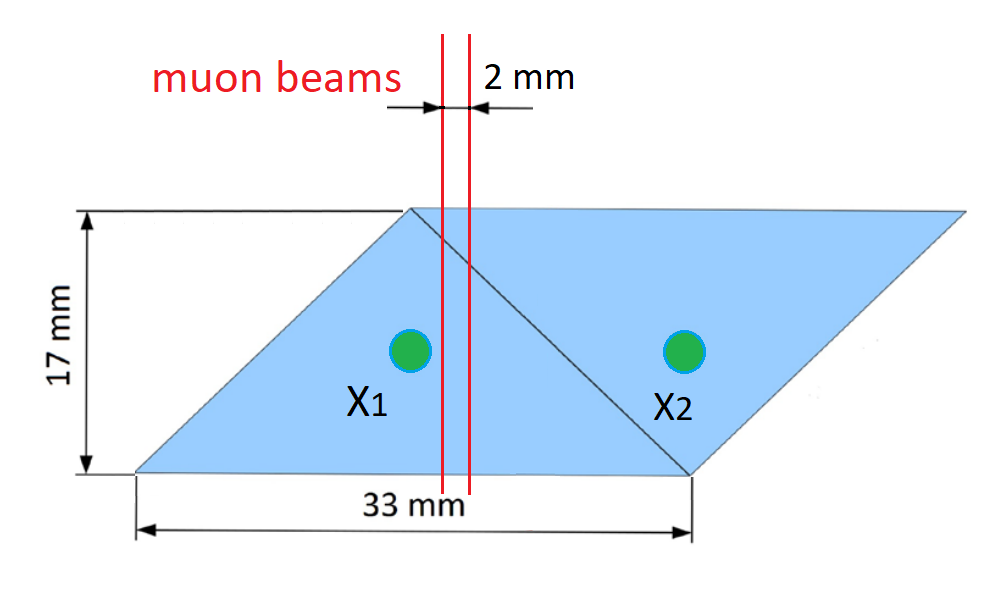}
\includegraphics[width=0.52\textwidth]{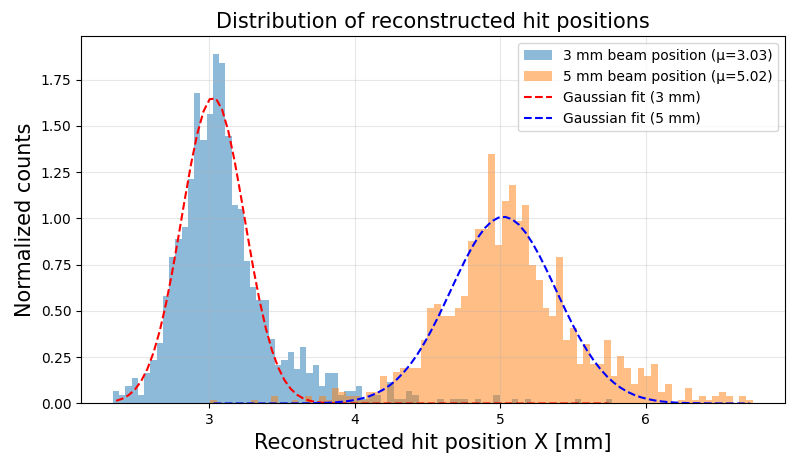}
\caption{(a) Cross-sectional view of the triangular scintillator bars illustrating the geometry used for hit position reconstruction. The red lines, separated by 2~mm, represent two incident muon trajectories simulated to evaluate the detector’s spatial resolution. 
(b) Distribution of reconstructed hit positions using asymmetry between upper and lower triangular scintillation bars for two simulated muon beams separated by 2 mm. }
\label{fig:f1reco}
\end{figure}

To evaluate the hit position, muon beams were simulated with a 2~mm step from position \(X_1\) to \(X_2\). For each beam position, 1000 muons were generated. The difference in the number of detected optical photons between the upper and lower triangular scintillation bars provides sensitivity to the muon impact position. %Figure~\ref{fig:f1reco}(a) shows the stacked triangular bars with muon trajectories separated by 2 mm. 
When two scintillation bars record signals, the muon impact position along the detector axis can be estimated from the ratio of detected photons, $N_2/N_1$, in the upper and lower bars. A geometric correction factor accounting for the track inclination angle $\vartheta$ is applied~\cite{anghel2015plastic}:
\begin{equation}
x = x_1 + \frac{a \, N_2}{N_1 + N_2} + h \frac{N_2}{N_1 + N_2} \tan \vartheta
\label{eq:impact_position}
\end{equation}
where $a$ is the fiber spacing ($a$ = $X_1$-$X_2$), $h$ the rod height, and $N_1$, $N_2$ the detected photons in upper and lower super layers. 
Figure~\ref{fig:f1reco}(b) shows the distribution of reconstructed hit positions for two muon beams separated by 2~mm. 
The two peaks correspond to beam positions at 3 and 5 mm (assuming \textit{X}$_1$~=~0), demonstrating that the light-yield asymmetry between the upper and lower triangular scintillation bars provides a sensitive measure of the muon impact point.
%h!
The number of optical photons detected in each layer was used to calculate the hit position according to Equation~\ref{eq:impact_position}. 
The position resolution, $\sigma_x$, can be estimated from the overlap region of the two distributions by fitting them with Gaussian functions and evaluating their standard deviations. 
This method enables the use of data from a pair of orthogonal layers (a superlayer) to determine the local hit position in both \textit{x} and \textit{y} coordinates.

\begin{figure}[h!]
\centering
\includegraphics[width=0.46\textwidth]{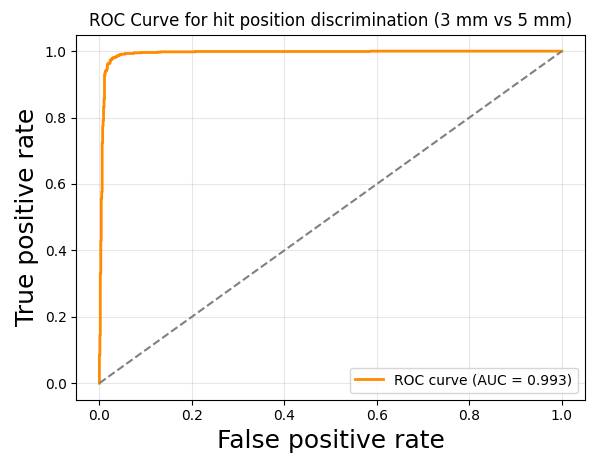}
\caption{The ROC curve for the distribution of reconstructed hit positions, obtained from the asymmetry between the upper and lower triangular scintillator bars for two muon trajectories separated by 2~mm.}
\label{fig:fROC}
\end{figure}

The ability of the reconstruction algorithm to distinguish closely spaced muon trajectories was evaluated using a Receiver Operating Characteristic (ROC) analysis. The ROC curve was generated from the distribution of reconstructed hit positions, obtained from the asymmetry between the upper and lower triangular scintillator bars, for two simulated muon beams separated by 2~mm. The resulting area under the curve (AUC) of 0.993, as shown in Figure~\ref{fig:fROC}, indicates excellent discrimination capability, confirming that the detector geometry and reconstruction method provide sufficient spatial resolution to reliably separate muon tracks with millimeter-scale spacing.

\subsubsection{Muon track reconstruction} 

\begin{figure}[h!]
\centering
\includegraphics[width=0.5\textwidth]{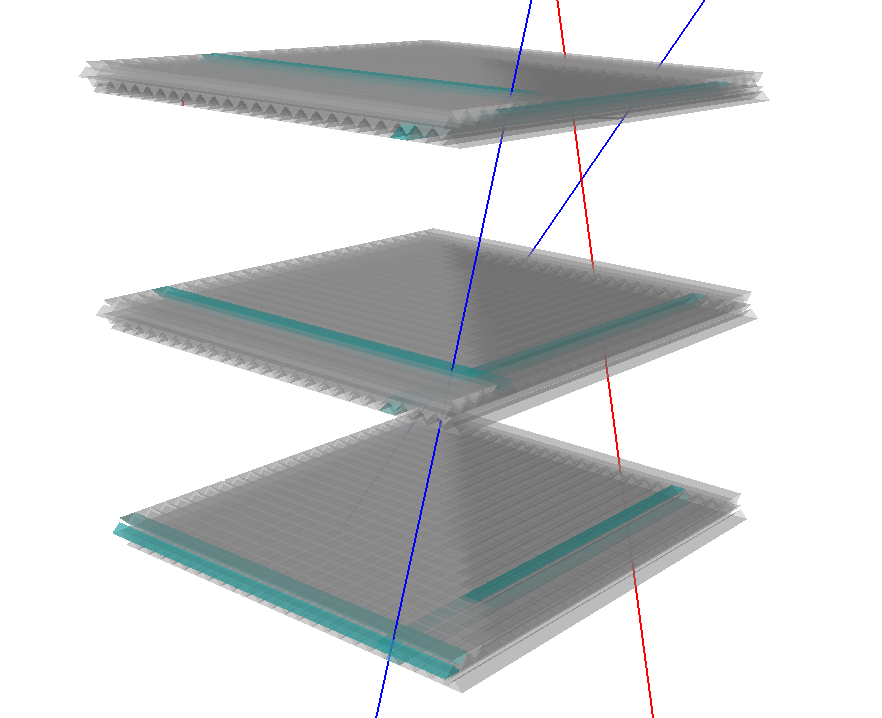}
\caption{Illustration of the muon tracking principle in a telescope, from GEANT4 simulation.
Muons with positive charge are shown in red and muons with negative charge in blue. A muon track is reconstructed when a spatial coincidence of hits is observed in all three tracking planes. Scintillation bars activated by the passing muon are highlighted in cyan. }
\label{fig:detector}
\end{figure}

The muon trajectory is discretized as it passes through scintillation layers, with the impact point called an \textit{hit position}. 
The telescope has three detection planes. Each plane comprises two orthogonal superlayers; each superlayer consists of triangular scintillator bars (strips) (as shown in Figure~\ref{fig:detector}). A muon crossing a plane typically deposits energy in several neighbouring triangular strips of a superlayer. For the geometry described, a muon typically produces up to four fired strips per plane (two in each orthogonal layer), so that across the three planes 12 strip register signals.
In realistic simulations, optical photon tracking for such a detector geometry would require simulating tens of thousands of photons per muon, a simplified approach is adopted. Each triangular scintillator bar is defined as a sensitive detector, and the energy deposition and hit position coordinates are recorded for every muon that passes through it.
Instead of using the number of optical photons detected, the energy deposition in each layer was used to calculate the hit position according to Equation~\ref{eq:impact_position}. This method allows data from a pair of orthogonal layers (a \textit{superlayer}) to determine the local hit position in both $x$ and $y$ coordinates with high precision.

In this approach, the degradation of position resolution due to Landau fluctuations in the $dE/dx$ energy loss, Poisson fluctuations in light collection (and photoelectron generation), and spatially dependent non-uniformities in light collection is not explicitly included. To account for these effects, the hit coordinates $(x, y)$ in each detection plane are smeared using a Gaussian distribution with a standard deviation chosen to reproduce a realistic spatial resolution of 2~mm. The final muon trajectory is then reconstructed by fitting these smeared hit positions across the three detector modules (superlayers).

\subsubsection{Muons event generator for modeling muon transport underwater}

To optimize computational resources, we adopted a strategy similar to that used for modeling underground muon fluxes, where the MUSIC (MUon SImulation Code) code is employed to calculate muon transport through rock and the resulting distributions are then used by MUSUN (MUon Simulations UNderground) to simulate muon flux in the underground laboratory and detector interactions~\cite{kudryavtsev2009muon}. MUSIC and MUSUN are essential tools for simulating cosmic-ray muons in underground research. MUSIC acts as a muon propagator, modeling energy loss and scattering as muons traverse rock or water overburden. It provides detailed energy and angular distributions of muons at various depths. MUSUN uses MUSIC’s output to generate realistic muon populations, defining their position, energy, and direction, as they would appear in or around an underground detector. Together, MUSIC and MUSUN enable accurate predictions of underground muon fluxes, crucial for experiment design and data interpretation in particle and astroparticle physics. By integrating accurate muon transport modeling with realistic muon generation, these tools enable the mitigation of muon-induced backgrounds, improving the sensitivity of dark matter, neutrinoless double beta decay, and other rare event searches.
Alternatively, the dedicated Monte Carlo codes (MUon inTensity codE) and PROPOSAL~\cite{woodley2024cosmic}, MUPAGE (MUon GEnerator from PArametric formulas)~\cite{carminati2008atmospheric} simulate muon transport and generate muons at underground sites. 

In our approach, the GEANT4 package was used instead of MUSIC to simulate muon propagation through water, from the sea surface down to the shipwreck location. Muons were generated at sea level with the CRY event generator~\cite{hagmann2007cosmic}, sampled over a 50~$\times$~ 50\,m\textsuperscript{2} surface.
Muons were propagated through 35 meters of water and interacted with a sensitive surface measuring 20 $\times$ 20 m$^2$. Out of 10$^9$ generated muons, approximately 1.17 $\cdot$ 10$^7$ reached this surface. 

For these muons essential information was recorded: particle type (PDG code), kinetic energy, interaction position, and muon momentum direction cosines ($u_x$, $u_y$, $u_z$).
Similar to the MUSUN code, the tabulated underwater muon energy and angular distributions were implemented in the \texttt{PrimaryGeneratorAction} class in GEANT4, allowing muons to be generated directly at a depth of 35 meters.
In GEANT4, the \texttt{PrimaryGeneratorAction} class (or a class derived from \texttt{G4VUserPrimaryGeneratorAction}) is a user-defined class that defines how primary particles are generated for a GEANT4 simulation. It is one of the three mandatory classes required for a GEANT4 application and is responsible for creating primary events by setting up particles, their initial positions, energies, and directions. 
For simple distributions, GEANT4 provides the capability to define particle generators based on mathematical, analytical formulas. For more complex or non-standard distributions, such as those found in real experimental data, it is possible to load data from external files.
This involves reading distribution information from data tables, which can then be used to generate primary particles with the desired physical properties.
For example, in the GEANT4 generic examples, the \texttt{FCALPrimaryGeneratorAction} class implements a particle generator that reads particle data from external files.
In our approach, the external file contains data on muon charge, position, energy, and angular distributions at a depth of 35 m. These data are used to generate muons at this depth, serving as primary particles in various filling scenarios to model their interactions with a parameterized shipwreck geometry.

\subsection{Muon imaging of the underwater objects}

\begin{figure}[h!]
\centering
\includegraphics[width=0.6\textwidth]{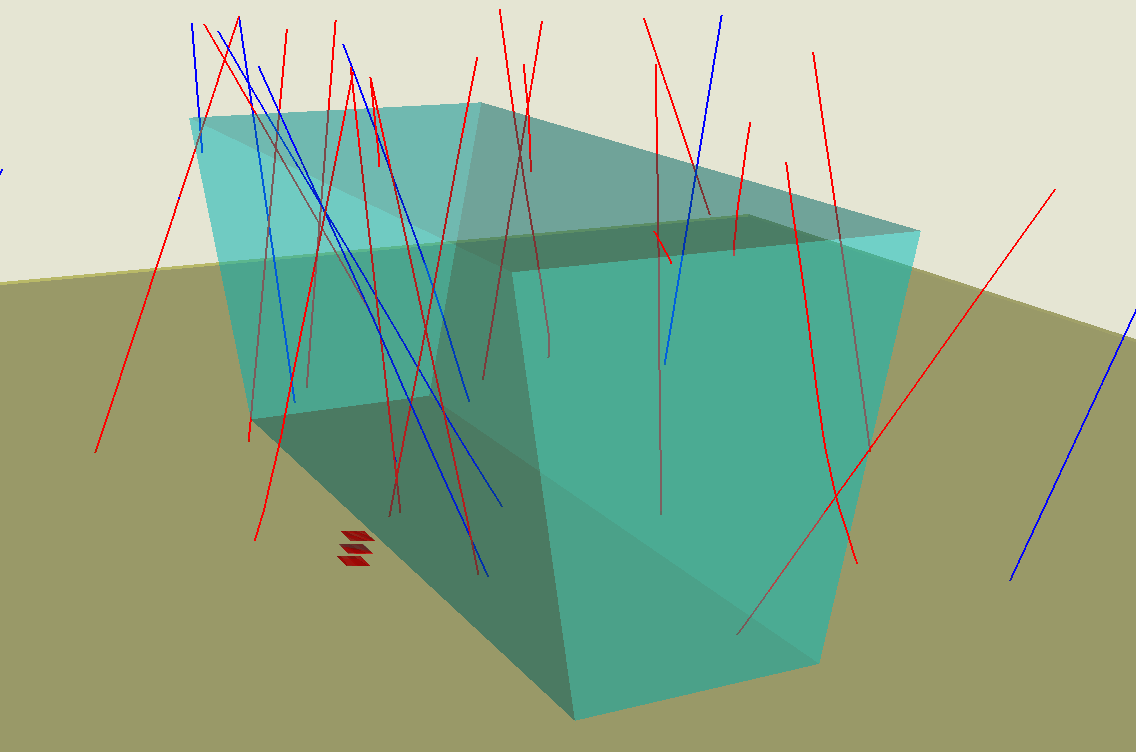}
\caption{Schematic of muography imaging of an underwater shipwreck, modeled as a trapezoidal structure, with a muon detector placed on the seabed and oriented toward the shipwreck. }
\label{fig:fship}
\end{figure}

In this study, we simulated a shipwreck as a trapezoidal object approximately 30 meters long, 15 meters high, and 15 meters wide, submerged at a depth of 50 meters to mimic the marine environment (as shown in Figure~\ref{fig:fship}). The wreck was filled with crude oil (density = 0.92 g/cm\textsuperscript{3}) or explosive hexogen (RDX) (density = 1.82 g/cm\textsuperscript{3}), depending on the scenario.

To analyze the contents of shipwrecks, such as determining whether they are filled with water, hazardous crude oil, or explosives, the transport of muon flux through the target can be simulated. Underwater muons were generated using the \texttt{PrimaryGeneratorAction class}, which utilized tabulated muon data at a depth of 35~m and sampled over a 20 $\times$ 20~m$^2$ surface. To achieve sufficient statistical accuracy, a total of 8~$\cdot~10^9$ muons were generated, corresponding to a scanning time of approximately seven days.
By examining the attenuation of muon flux as it passes through different materials, the muon transmission (MT) technique enables the creation of two- and three-dimensional average transmission and density maps of the studied object, providing valuable insights into internal relative densities.  Achieving a well-defined muographic image of large targets, such as underwater shipwrecks, often requires data acquisition periods extending to weeks or even months. This extended duration is necessary to accumulate sufficient muon statistics across various viewing angles, ensuring adequate resolution and contrast in the resulting density maps. 
The MT technique involves two measurement configurations: with target and free-water (no target). The target configuration involves measuring the muon flux near or beneath the object of interest, where the water or material thickness between the detector and the surface constitutes the target. 
Angular fluxes are computed from reconstructed muon tracks, implicitly including
acceptance and efficiency via the identical selection. The simulated
transmission for a given column density $X$ along each line of sight is
\begin{equation}
t(\varphi,\theta; X) =
\frac{\Phi_{{target}}(\varphi,\theta; X)}
     {\Phi_{{notarget}}(\varphi,\theta)}
\end{equation}
where no-target (free-water) configuration consists of ship filled with water extending to a depth of \textit{D} meters, while the target configuration includes a ship filled with either crude oil or RDX. The result is a lookup table that maps column density \( X \) (in g/cm$^2$) to the corresponding simulated transmission \( t(\varphi,\theta; X) \) for each angular bin.
Finally, the measured transmission values are associated with their corresponding lines of sight and projected onto a 2D image plane, yielding a two-dimensional density map of the shipwreck. In practice, the detector is positioned on the seabed, oriented toward the target, and muon counts $N_{\mathrm{target}}(\varphi,\theta)$ are accumulated over the acquisition time $t_{\mathrm{acq}}$ for each directional bin.

In the inversion step, the measured transmission 
\( t_{\mathrm{meas}}(\varphi, \theta) \) 
is compared with the simulated transmission 
\( t_{\mathrm{simu}}(\varphi, \theta; X) \) 
to determine the reconstructed (inverted) column density 
\( \hat{X}(\varphi, \theta) \).
Subtracting the water baseline yields the excess or deficit in column density attributable to the shipwreck's internal structure.
For each angular bin $(\varphi,\theta)$, solve
\[
t_{\text{simu}}(\varphi,\theta;\hat{X}) \approx t_{\text{meas}}(\varphi,\theta)
\]
for $\hat{X}(\varphi,\theta)$. The uncertainty is
\[
\delta X(\varphi,\theta) \approx 
\frac{\delta t_{\mathrm{meas}}}
{\left|\partial t_{\text{simu}}/\partial X\right|}.
\]
Subtract the water baseline:
\[
\Delta \hat{X}(\varphi,\theta) = \hat{X}(\varphi,\theta) - X_{\text{water}}(\varphi,\theta).
\]
To deriving material density we assume knowledge of the chord length through the wreck $L_{wreck}(\varphi,\theta)$, obtained via sonar scans or hull models, the average material density along each line of sight is:
\begin{equation}
\bar{\rho}(\varphi,\theta) = \frac{\Delta \hat{X}(\varphi,\theta)}{L_{\text{\textit{wreck}}}(\varphi,\theta)}.
\end{equation}

The detector field of view is discretized into angular bins, indexed by $k$ and $l$, corresponding respectively to the azimuthal angle $\varphi_k$ and zenith angle $\theta_l$. For each angular bin $(\varphi_k, \theta_l)$, the reconstructed muon direction $\hat{\mathbf{r}}(\varphi_k, \theta_l)$ defines a ray originating from the detector center $\mathbf{o}$. Its intersection with the target (projection) plane, located at distance $R$, is given by
\[
\mathbf{p}_{kl} = \mathbf{o} + R\,\hat{\mathbf{r}}(\varphi_k,\theta_l).
\]

The spatial resolution depends on the detector's angular resolution $\sigma_{det}$ and the multiple scattering angle $\theta_0$, which broadens the point spread function (PSF):
\begin{equation}
\sigma_{\text{\textit{psf}}} \approx R \times \sqrt{\sigma_{det}^2 + \theta_0^2},
\end{equation}
with~\cite{pdg2020}
\begin{equation}
\theta_0 \approx \frac{13.6\,\text{MeV}}{\beta p c} \sqrt{\frac{x}{X_0}} \left[ 1 + 0.038 \ln \left( \frac{x}{X_0} \right) \right],
\end{equation}
where $x$ is the material thickness traversed, $X_0$ is the radiation length, \textit{c} is the speed of light, $\beta$ is the ratio of speed of the muon to speed of light and \textit{p} is the momentum of the muon (in MeV/c).

\section{Results}

\begin{figure}[t]
\centering
\includegraphics[width=0.65\textwidth]{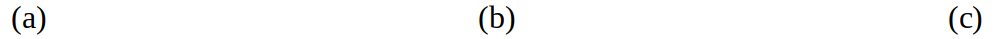}
\includegraphics[width=0.97\textwidth]{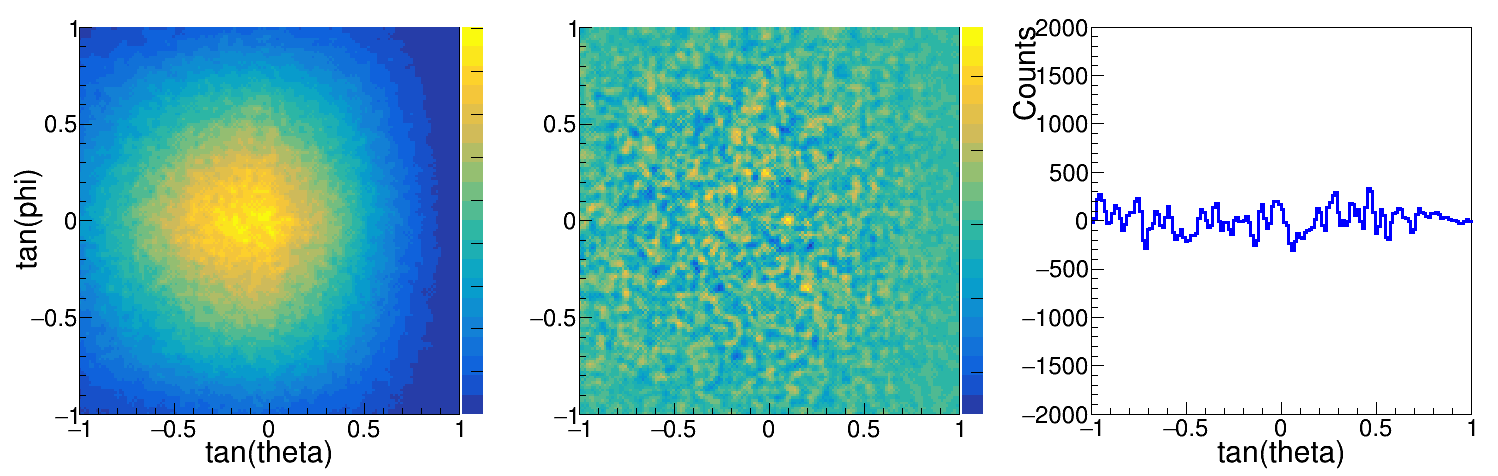}
%\vspace{-3mm}
\caption{(a) Angular distribution of muon counts simulated for a shipwreck completely filled with water, with the detector at a depth of 50 m. (b) Difference in angular distributions of muons between two configurations, both with the shipwreck filled with water. (c) Projection of the 2D difference histogram in panel (b) onto the x-axis.
The numerical values of the color maps are the same as those on the projection histogram in panel.}
\label{fig:f2}
\end{figure}

\begin{figure}[t]
\centering
\includegraphics[width=0.65\textwidth]{figures/abc.png}
\includegraphics[width=0.97\textwidth]{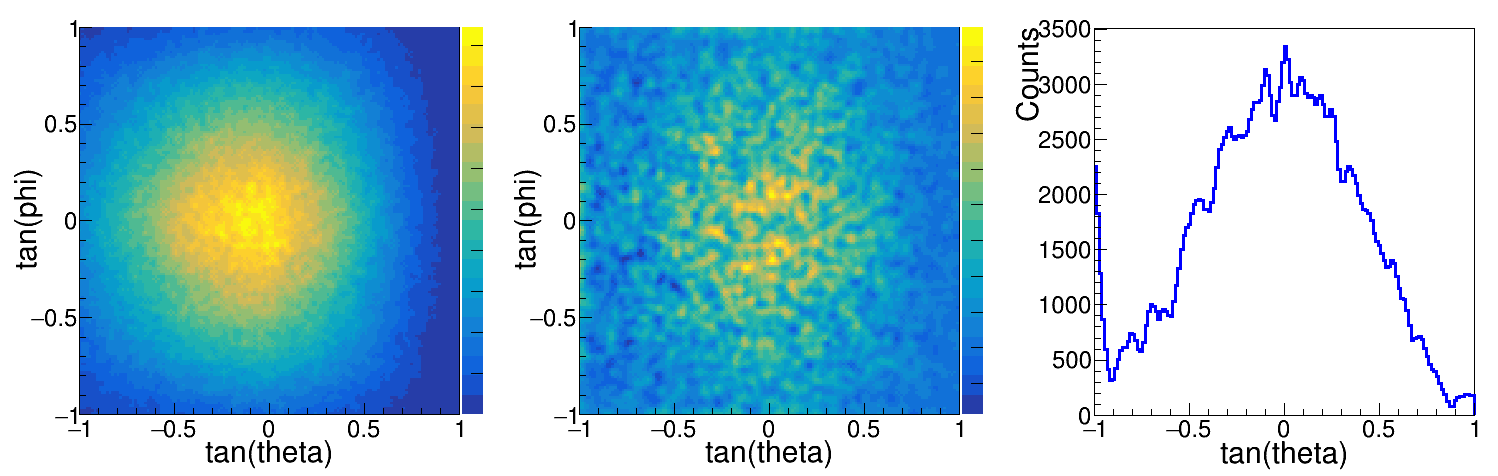}
\caption{(a) Angular distribution of muon counts simulated for a shipwreck filled with crude oil, with the detector at a depth of 50 m. (b) Difference in angular distributions of muons between the water-filled and crude oil–filled configurations. (c) Projection of the 2D difference histogram in panel (b) onto the x-axis. 
}
\label{fig:f3}
%\end{figure} 
%\begin{figure}[t]
\centering
\includegraphics[width=0.65\textwidth]{figures/abc.png}
\includegraphics[width=0.97\textwidth]{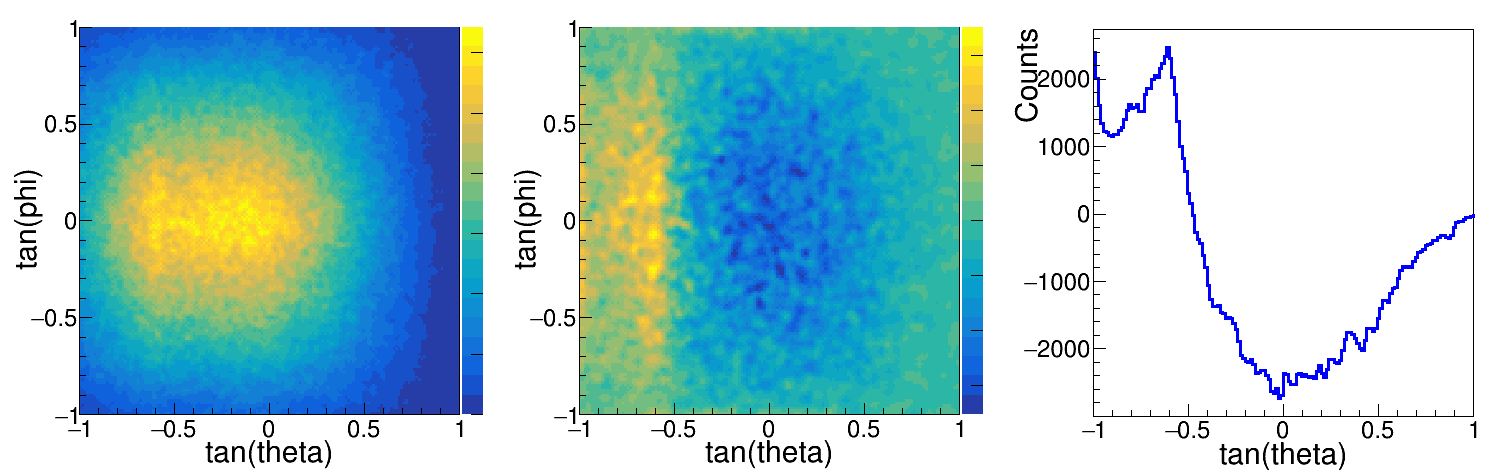}
\caption{(a) Angular distribution of muon counts simulated for a shipwreck filled with water. (b) Difference in angular distributions of muons between the water-filled and explosive-filled configurations. (c) Projection of the 2D difference histogram in panel (b) onto the x-axis. 
}
\label{fig:f4}
\end{figure}

The simulated two-dimensional transmission maps for different scenarios, a shipwreck filled with water, crude oil, and RDX, are presented in Figures~\ref{fig:f2}--\ref{fig:f4}, assuming an exposure time of one week. Interpolation between angular bins is applied to produce a smooth density map. 

Figure~\ref{fig:f2}(a) shows the angular distribution of muon counts simulated in the free-water configuration, with the shipwreck filled with water and the detector deployed at an underwater depth of 50 m. Figure~\ref{fig:f2}(b) presents the difference in angular distribution of muons between two configurations where the shipwreck is filled with water in both cases. Figure~\ref{fig:f2}(c) displays the one-dimensional projection of the difference histograms between these water-filled configurations. Figures~\ref{fig:f2}(b) and (c) exhibit a uniform distribution, as both images correspond to the same configuration of the shipwreck filled with water, resulting in negligible differences.
Figure~\ref{fig:f3}(a) shows the angular distribution of muon counts simulated in the water-target configuration, with the shipwreck filled with crude oil. Figure~\ref{fig:f3}(b) presents the difference (subtraction) of the angular distribution of muons between the water and crude oil configurations. Figure~\ref{fig:f3}(c) displays the one-dimensional projection of the difference histograms between the water and crude oil configurations, highlighting the muon excess corresponding to the crude oil-filled region due to less absorption from its lower density.
Figure~\ref{fig:f4}(a) shows the angular distribution of muon counts simulated in the water-target configuration, with the shipwreck filled with RDX. Figure~\ref{fig:f4}(b) presents the difference (subtraction) of the angular distribution of muons between the water and RDX configurations. Figure~\ref{fig:f4}(c) displays the one-dimensional projection of the difference histograms between the water and RDX configurations, highlighting the muon deficit corresponding to the RDX-filled region due to greater absorption from its higher density. The significant deficit in muon flux observed in the x-projection of the difference histograms (water - RDX configurations, Figure~\ref{fig:f4}(c)) at \(\tan(\theta)\) values of -0.5 -- 0.5 can be attributed to the higher density of RDX compared to sea water, resulting in reduced muon flux.

%
%\vspace{-5mm}

\section{Discussion}

With sufficient angular coverage and extended exposure times, combining 2D projections from multiple viewing positions enables the reconstruction of comprehensive three-dimensional tomographic models. 
This multi-angle approach facilitates systematic analysis of muon attenuation, scattering, and transmission properties, allowing for the characterization of the wreck’s internal structure with improved accuracy. 
When paired with detectors featuring high spatial and angular resolution, this configuration can produce detailed 3D images of internal features and estimate material densities with greater precision. Similar methodologies have been successfully employed in terrestrial applications; for example, muon imaging has been used to investigate the internal structure of decommissioned nuclear reactors, demonstrating the technique’s capability for high-resolution 3D characterization \cite{procureur20233d}.  

For underwater imaging, the detector’s spatial resolution is a critical factor influencing the accuracy of 3D reconstruction. In this study, a muon telescope was constructed using triangular extrusion scintillator bars, providing a spatial resolution of 2–3 mm as a cost-effective solution. 
However, if higher resolution is required for detailed reconstruction of the shipwreck’s internal structure, more advanced muon tracking systems with resolutions of 0.5–1 mm, such as those discussed in~\cite{Georgadze2025,MARESEC}, can be employed.

It is important to note that this study did not account for seasonal variations in muon flux or changes in water column density, as the primary objective was to evaluate the fundamental feasibility of the method. Incorporating such environmental factors in future studies could improve the accuracy of density estimates and enhance the robustness of the technique. Although these simplifications may introduce minor uncertainties in the reconstructed density values, they are unlikely to affect the overall assessment of the method’s potential.
Muon generation in this work was performed using the CRY event generator, which tends to produce fewer muons (approximately 7–8×10\(^3\) muons per m\(^2\) per minute) than the commonly referenced value of around 10,000 muons per m\(^2\) per minute. Consequently, the estimates presented here should be regarded as conservative. Additionally, CRY allows for adjustments based on geographic location, including the geomagnetic cutoff and solar cycle effects, ensuring that simulations can be tailored to specific sites. 

Additionally, the thickness of the shipwreck hull was not simulated due to simplifications and computational limitations. This omission is justified, as 2 cm of steel corresponds to approximately 15.6 cm of water in terms of muon energy loss. Compared with the $\approx$15 m thickness of water or oil considered in the model, this introduces only a small systematic uncertainty (1-2\%) in the muon flux, primarily affecting lower-energy muons, without significantly altering the observed differences between the water- and oil-filled configurations.

\section{Conclusion}

This study evaluated the feasibility of muon radiography as a non-invasive technique for detecting and characterizing submerged structures and hazardous materials within shipwrecks. Using Monte Carlo simulations, we modeled a simplified scenario of muon imaging for a trapezoidal shipwreck approximately 30 m long, 15 m high, and 15 m wide, located at a depth of 50 m. Three internal filling conditions were considered: water, crude oil, and the high-density explosive RDX. Assuming a one-week exposure time, the simulation results indicate that muon imaging can resolve density contrasts sufficient to distinguish between these materials, demonstrating the method’s potential as a practical tool for underwater hazard assessment. 

These preliminary findings confirm the feasibility and promising potential of muon radiography for non-invasive identification and characterization of shipwreck contents in challenging underwater environments such as the Baltic Sea, while acknowledging that further refinements, such as accounting for environmental variability and optimizing detector performance, are essential for advancing practical applications.

Beyond environmental and safety applications, this approach offers promising opportunities in marine archaeology, geology, and underwater cultural heritage research. By enabling detailed, non-destructive imaging of submerged structures and geological formations, muon radiography can enhance our understanding of underwater sites, support preservation efforts, and reveal historical and geological insights that are otherwise difficult to obtain.

\acknowledgments
This research did not receive any funding.

%\paragraph{Note added.} This is also a good position for notes added after the paper has been written.

\bibliographystyle{JHEP}
\bibliography{bibliography}
\end{document}